\documentclass[12 pt, a4paper]{article}
\usepackage{physics}
\usepackage{jheppub}
\usepackage{bm}
\usepackage{xcolor}
\usepackage{amsmath}
\usepackage{amsfonts}
\usepackage{url}
\usepackage{graphicx}
\usepackage{ragged2e}
\usepackage{caption}
\usepackage{subcaption}
\usepackage{chngcntr}
\counterwithout{equation}{section}

\usepackage{lmodern}
\usepackage[utf8]{inputenc}
\usepackage[T1]{fontenc}
\definecolor{refkey}{gray}{0.45}
\definecolor{labelkey}{RGB}{155,48,48}
\hypersetup{
  colorlinks=true,
  citecolor=magenta,
  linkcolor=blue,
  urlcolor=violet,
  pdftitle={Upamanyu Moitra: Lumos Extrema},
  pdfauthor={Upamanyu Moitra},
 }

\newcommand{\mi}{\mathrm{i}}

\newcommand{\te}{\theta}

\newcommand{\om}{\omega}

\newcommand{\la}{\lambda}
\newcommand{\La}{\Lambda}
\newcommand{\e}{\epsilon}

\newcommand{\bee}{\begin{equation}}
\newcommand{\ee}{\end{equation}}
\newcommand{\beq}{\begin{equation*}}
\newcommand{\eeq}{\end{equation*}}
\newcommand{\baa}{\begin{equation}\begin{aligned}}
\newcommand{\ea}{\end{aligned}\end{equation}}
 
\newcommand{\qh}{\hat{q}}

\newcommand{\kh}{\hat{k}}
\newcommand{\oh}{\hat{\omega}}

\title{\center{ \fontfamily{lmr}\selectfont Lumos Extrema}}

\author{\center \fontfamily{lmr}\selectfont Upamanyu Moitra\footnote{Present Address:  Institute of Physics,  Universiteit van Amsterdam,  Science Park 904, 1089 XH Amsterdam,  The Netherlands.  Email: \href{mailto:u.moitra@uva.nl}{\texttt{u.moitra@uva.nl}}}}

\affiliation{\begin{center}The Abdus Salam International Centre for Theoretical Physics (ICTP)\\
Strada Costiera 11,  Trieste 34151,  Italy

\href{mailto:umoitra@ictp.it}{\texttt{umoitra@ictp.it}}
\end{center}}

\abstract{{\centering
\begin{justify}
We consider trajectories of massless particles in the presence of charged black holes in asymptotically AdS spacetimes in arbitrary dimensions.  We study the properties of the photon ring in the (near-)extremal limit and show that the photon ring can probe the near-horizon region in two different scenarios: in high enough number of spacetime dimensions or when the massless particle carries an electric charge.  We propose a simple $\mathrm{i} \epsilon$-prescription for implementing the JWKB approximation and show its utility in various contexts.  We calculate the quasi-normal modes for charged fields in the eikonal limit and show the emergence of a new time-scale of relaxation in the dual field theory side.  In the near-extremal limit, we show generally that the time-scale of decay of the perturbations is governed by the chemical potential.  We also verify our analytical results numerically.   The $\mathrm{i} \epsilon$-prescription allows us to study the superradiant modes and the associated instabilities in the eikonal limit easily.  We comment on some related aspects regarding near-extremal black holes.
\end{justify}
}

}

\dedicated{To my motherland, India,  in gratitude \ldots}

\makeatletter
\gdef\@fpheader{}
\makeatother

\begin{document}
\maketitle

\section{Introduction}\label{sec-intro}

The phenomenon of massive bodies influencing the trajectory of massless particles was well appreciated even before the general theory of relativity was completely formulated.   A proper relativistic treatment not only gives the correct value for the angle of deflection of light by massive objects \cite{Einstein:1916}, it makes a novel prediction \cite{Hilbert1917} about the existence of \emph{unstable} circular photon orbits in the presence of compact objects such as black holes.   These orbits are known as photon rings or photon spheres,  from which massless particles either escape to asymptotic infinity or fall into the black hole.  In this sense,  the photon ring can be thought to lie at the edge of the region of the geometry accessible to an external observer.  The properties of photon rings have been extensively studied and it has also been argued that their existence is concomitant with event horizons of astrophysical black holes \cite{Cardoso:2014sna}. The notion of photon surfaces has been generalized to different spacetime geometries \cite{Claudel:2000yi}.

Recent progress in directly imaging black holes \cite{EventHorizonTelescope:2019dse} has raised the possibility that signatures of the black hole photon ring could be directly observed \cite{Johnson:2019ljv}.  Besides their astrophysical significance,  black holes are indispensable tools in furthering our understanding of quantum gravity.   The most well-understood class of black holes in this context are \emph{extremal},  containing the maximum amount of charge and/or spin for a given mass.  The Bekenstein-Hawking entropy for certain extremal black holes has been reproduced by microscopic calculations in string/M-theory -- first for asymptotically flat black holes \cite{Strominger:1996sh} and, relatively recently, for black holes in asymptotically Anti-de Sitter (AdS) spacetime as well (see \cite{Zaffaroni:2019dhb} for a review).  It turns out that one is able to deduce universal features even about black holes which are slightly away from the extremal limit.  It was established in \cite{Moitra:2019bub, Moitra:2022ooc}  that the two-dimensional Jackiw-Teitelboim (JT) model \cite{Teitelboim:1983ux,  Jackiw:1984je} arises in the near-horizon description of near-extremal black holes in a universal way in any general supergravity theory with arbitrary charges,  rotations and matter content.  One can thus make universal statements about the dynamics and thermodynamics of the black holes essentially working with the Schwarzian mode associated with the JT model  possibly augmenting it with other ```phase modes'' if necessary \cite{Moitra:2018jqs}.  It is worth mentioning in this regard that that near-extremal black holes are not merely an academic curiosity --- it is believed that highly spinning near-extremal black holes would be present in nature as well. In this article, with a view towards understanding (near-)extremal black holes,  we set our focus on charged black hole backgrounds in an asymptotically AdS spacetime.  

Important information about a geometry can be gleaned from a study of quasi-normal modes (QNMs) \cite{Berti:2009kk,  Konoplya:2011qq}, governing the behavior of linear perturbations.  In the context of the AdS/CFT correspondence \cite{Maldacena:1997re, Gubser:1998bc,  Witten:1998qj},  QNMs are associated with thermalization \cite{Horowitz:1999jd} in the dual conformal field theory (CFT) and correspond to the  poles of the retarded Green functions \cite{Son:2002sd}.  We are therefore able to learn about certain properties of strongly coupled field theory from the quasi-normal spectrum in the weakly coupled gravity theory.  In appropriate limits,  the QNM spectrum is closely tied to particle trajectories in the bulk gravitational theory \cite{Festuccia:2008zx}.   Recent studies of bulk particle trajectories and their translation into the CFT language have shed light on various aspects of AdS/CFT \cite{Hashimoto:2019jmw, Berenstein:2020vlp,  Liu:2022cev,  Dodelson:2022eiz, Moitra:2023yyc, Hashimoto:2023buz,  Dodelson:2023nnr}.

QNMs corresponding to the photon sphere in an asymptotically flat geometry were first considered in \cite{Cardoso:2008bp}.  Several interesting aspects of the photon sphere in the asymptotically flat context were pointed out in \cite{Raffaelli:2021gzh, Hadar:2022xag, Kapec:2022dvc}. Such considerations have recently been extended to uncharged,  non-rotating black holes in AdS in \cite{Hashimoto:2023buz,  Dodelson:2023nnr}.  In this article,  we study massless particle trajectories for charged, spherically (or planar) symmetric black holes in an arbitrary-dimensional asymptotically AdS spacetime with particular interest in the (near-)extremal situation  --- the flat limit, when it exists,  is easy to take. In the presence of a conserved global charge in a CFT,  one has in addition to temperature another dimensionful scale -- the chemical potential.  In the near-extremal limit,  when we take the temperature to be very small,   relaxation time-scales are governed by the chemical potential and not by the temperature.  This happens, for example,  for the low-lying modes analogous to hydrodynamic modes near extremality \cite{Moitra:2020dal} even when the conventional hydrodynamic description itself breaks down. In this article, we investigate the eikonal spectrum,  far removed from the hydrodynamic regime.  The aforementioned general physical intuition tells us that the chemical potential should still determine the quasi-normal spectrum.  One aim of the article is to establish the validity of this intuition explicitly.

A key goal of this study is to consider to what extent the photon ring could probe near-horizon physics.  As we shall see,  photon rings are generally quite far from the horizon.  In fact,  even at extremality, they are often located in a region where the JT gravity description is no longer valid.  However,  we observe no fewer than two scenarios in which the photon ring can be brought very close to the horizon --- first, when the number of spacetime dimensions becomes large. A large number of spacetime dimensions is often a good expansion parameter to describe gravitational dynamics --- see \cite{Emparan:2020inr} for a review --- which emerges naturally in this context.  Another method of probing the horizon is to consider massless particles which are \emph{electrically charged}. In this case,  we find that the ``charged photon ring'' can be placed at any position of our choice.  We will argue that such charged rings can arise in several realistic situations, including dimensional reduction of rotating black holes.  { It is worth mentioning that near-horizon light rings have previously been considered in the context of rotating black holes \cite{Bardeen:1973tla,  Gralla:2017ufe,  Gates:2020els}.}

The authors of \cite{Festuccia:2008zx} had outlined a general procedure for finding QNMs by means of the Jeffreys-Wentzel-Kramers-Brillouin (JWKB) approximation\footnote{Given that the connection formulas -- which are central to our discussion -- were first discovered by Jeffreys,  we choose to err on the side of inclusion in calling it the JWKB method.}.   In this work, we propose a relatively simple $\mi \epsilon$-prescription for performing this approximation,  supplemented with a careful treatment of asymptotic expansions \cite{Dingle:1973}. When we can compare,  our results agree exactly with the complex JWKB method described in \cite{Festuccia:2008zx}. For a massless uncharged probe in the black brane spacetime, we confirm the intuitive expectation that the decay time-scale is indeed governed by the chemical potential,  with the same scaling as the temperature as observed in \cite{Festuccia:2008zx}.  We provide numerical evidence for this result, using an existing package \cite{Jansen:2017oag}; see also \cite{Morgan:2009vg} for related work in the uncharged case. When we consider massless charged probe particles,  we find a markedly different behavior for the quasi-normal mode, with a momentum scaling that is different from its uncharged cousin. Yet again,  we provide numerical evidence for this scaling.  The asymptotically globally AdS case can be treated along same lines --- we also make comments on photon sphere modes for complex charge and spin,  localized close to the horizon,  extending the results obtained in \cite{Dodelson:2023nnr}.

The $\mi \e$-prescription that we propose is easily amenable to the situation in which when one considers superradiant modes \cite{Brito:2015oca}. Superradiant instabilities in the eikonal limit, when present, and their associated time-scales are naturally obtained in this set-up. 

In section \ref{sec-phot}, we discuss in detail massless particle orbits,  both in the charged and uncharged cases.  In section \ref{sec-qnm},  we calculate the QNMs in several different scenarios though a detailed analysis.  Superradiant modes and  the associated instabilities are explored in section \ref{sec-superr}.  We conclude this paper with a discussion on some broader aspects of near-extremal black holes in \ref{sec-disc}.

\section{Photon Rings Near Near-Extremal Horizons}\label{sec-phot}

We consider the Einstein-Maxwell action in asymptotically AdS$_{d+1}$ spacetime,
\baa
S = \frac{1}{16 \pi G} \int \dd[d+1]{x} \sqrt{-g} \pqty{ R - F_{\mu \nu} F^{\mu \nu} + \frac{d(d-1)}{L^2}  }, \label{einmax}
\ea
where $L$ is the AdS radius. The electrically charged Reissner-Nordstr\"om solution in asymptotically globally AdS spacetime is given by the metric
\baa
\dd{s}^2 = - f(r) \dd{t}^2 + \frac{\dd{r}^2}{f(r)} + r^2 \dd{\Omega}_{d-1}^2 \label{metric},
\ea 
and the gauge field
\baa
A = \pqty{ \mu - c_d \frac{Q}{r^{d-2} } } \dd{t}, \label{gauge}
\ea
where $c_d = \sqrt{(d-1)/2(d-2)}$ and the red-shift factor $f(r)$ in the metric \eqref{metric} is given by
\baa
f(r) = 1 + \frac{r^2}{L^2} - \frac{2M}{r^{d-2} } + \frac{Q^2}{r^{2d-4}}.  \label{redshift}
\ea
For appropriate values of the parameters,  a black hole solution exists when for some positive value of $r  = r_+$,  $f(r_+)=0$ and $f(r) >0$ for all $r> r_+$.  In this article,  we work with a gauge in which the constant $\mu = c_d Q/r_+^{d-2}$ is identified with the chemical potential and the gauge potential \eqref{gauge} vanishes on the outer horizon. Note that in these units the chemical potential $\mu$ is dimensionless --- the dimensionful quantity, for relevant purposes, would be given by $\mu / L$.

In describing the exterior of the black hole,  we will sometimes find it convenient to present the results using the negative of the Regge-Wheeler coordinate instead of the radial coordinate $r$,
\baa
z = - r^* = \int_r^\infty \frac{\dd{r}}{f(r)}. \label{regwhee}
\ea
In this coordinate,  the asymptotic infinity at $r = \infty$ is mapped to $z = 0$ and the black hole horizon is mapped to $z = \infty$.

Let us consider the motion of massless particles in this geometry. For a null geodesic with the affine parameter $\lambda$, the quantities $E \equiv f(r) \dv*{t}{\lambda}$ and $\ell \equiv r^2 \dv*{\phi}{\lambda}$ will be conserved and the particle trajectory satisfies the effective one-dimensional equation
\baa
\pqty{ \dv{r}{\la} }^2 + \ell^2 \frac{f(r)}{r^2} - E^2 = 0.  \label{radeq}
\ea
Individually,  the values of $E$ and $\ell$ depend on a choice of affine parametrization -- the only physically meaningful quantity in this context is the ratio $\ell/E$,  which is related to the ``impact parameter''.

It is easy to see that circular photon orbits must satisfy
\baa
\eval{\dv{r} \pqty{ \frac{f(r)}{r^2} }}_{r =  r_p} = 0. \label{rpdef}
\ea
This immediately yields the closed-form solution
\baa
r_{p \pm} (M,  Q) = \bqty{ \frac{d M}{2} \pm \sqrt{\frac{d^2 M^2}{4} - (d-1) Q^2 }  }^{ \frac{1}{d-2} }.  \label{rppm}
\ea
There are two circular photon orbits. There is no explicit dependence of the radii on the AdS radius $L$.  We shall see shortly that only one of these -- the outer one -- is of physical interest to us.

One of our key objectives is to examine the system near the zero-temperature limit,  corresponding to near-extremal black holes.  In the extremal limit,  $f(r)$ would have a double zero at $r = r_h$, the extremal horizon.  The value of $M$ and $Q$ are parametrically related as \cite{Moitra:2019bub, Moitra:2022ooc}
\baa
M_{\rm ext} &= \pqty{ \frac{1}{r_h^2} + \frac{d-1}{d-2} \frac{1}{L^2} } r_h^{d}, \\
Q_{\rm ext} &= \pqty{ \frac{1}{r_h^2} + \frac{d}{d-2} \frac{1}{L^2} }^{1/2} r_h^{d-1}.  \label{extmq}
\ea
In this limit,  the near-horizon geometry factorizes into $\mathrm{AdS}_2 \times S^{d-1}$,  with the $\mathrm{AdS}_2$ radius given by
\baa
 L_2 = \pqty{ \frac{(d-2)^2}{r_h^2} + \frac{d(d-1)}{L^2}   }^{-\frac12}.  \label{l2def}
\ea
The factorized geometry is a good description for $(r-r_h) \ll r_h$, but beyond this limit, the geometry begins to exhibit departures from $\mathrm{AdS}_2 \times S^{d-1}$.

\begin{figure}
\centering
\includegraphics[width=12cm]{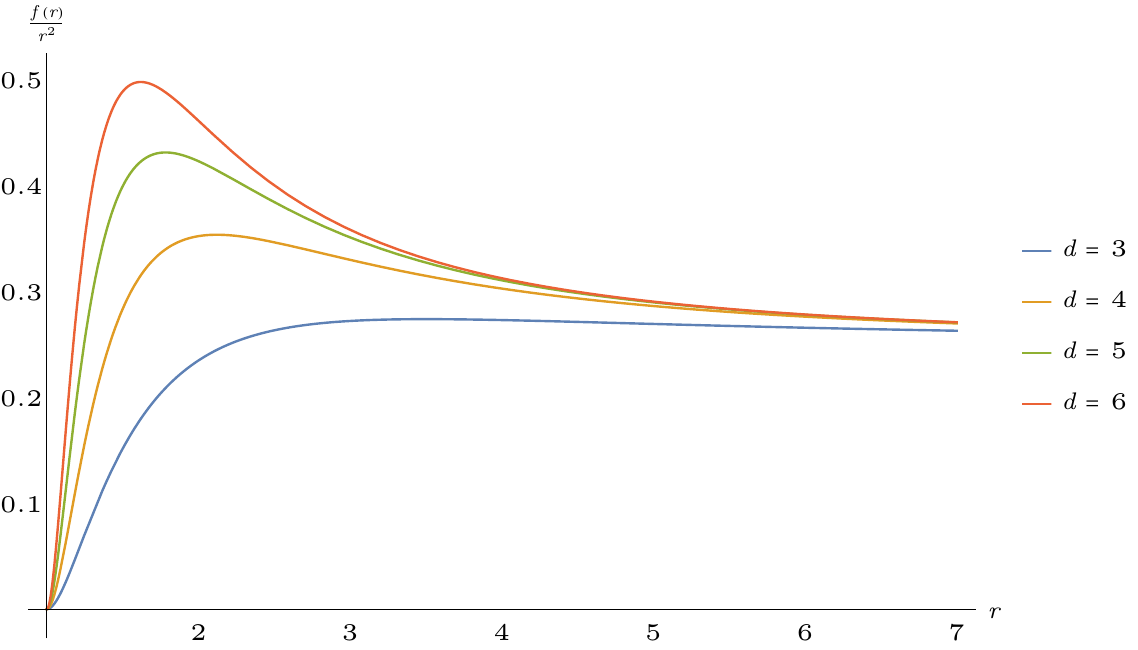}
\caption{Plots of the massless particle potential,  $f(r)/r^2$ for extremal black hole geometries for $d=3, 4, 5, 6$ for parameter values $r_h = 1,  L= 2$. The unstable photon sphere moves closer to the horizon at $r = 1$ for larger $d$.}
\label{fig-lardps}
\end{figure}

The branch at the smaller radius is exactly the extremal horizon
\baa
r_{p - } (M_{\rm ext},  Q_{\rm ext})  = r_h \label{innerphots}
\ea
for any value of $L$.  This branch is however not of interest to us since it goes behind the horizon as the temperature is increased.  The easiest way to see this is in the limit $Q \to 0$,  in which case $r_{p -} \to 0$.  We thus focus on the physically important outer branch $r_{p +}$.   This admits the exact expression
\baa
r_{p+} (M_{\rm ext},  Q_{\rm ext} ) = r_h (d-1)^{ \frac{1}{d-2} } \pqty{ 1 + \frac{d}{d-2}  \frac{r_h^2}{L^2} }^{\frac{1}{d-2}}.  \label{rphpl}
\ea
This corresponds to an unstable maximum of the photon potential,  as shown in Figure \ref{fig-lardps}  for various dimensions.  

For asymptotically flat extremal black holes in $d=3$, for example,  this photon orbit is located at $r = 2r_h$,  one horizon radius away from the horizon.  For uncharged black holes with horizon radius $r_0$,  it has been long known \cite{Hilbert1917} that the photon ring is located at $r = 3r_0$.  Thus, this photon orbit is not close to the horizon in any sense --- this orbit cannot probe the near-horizon $\mathrm{AdS}_2$ region in the extremal case.

The expression \eqref{rphpl} however naturally suggests a limit which can bring the photon ring close to the horizon --- the large dimension, $d \gg 1$ limit.  In taking the large $d$ limit,  we will always keep the horizon radius $r_h$ and the $(d+1)$-dimensional AdS radius $L$ to be fixed.  This will ensure,   among other features,  that the extremal chemical potential $\mu$,  defined previously, will have a finite limit as $d \to \infty$.  Keeping $r_h / L$ fixed,  we find that
\baa
r_{p +} (M_{\rm ext},  Q_{\rm ext}) = r_h \bqty{ 1 + \frac{\log d}{d}  + \frac{\log(1 + \frac{r_h^2}{L^2})}{d}  +  \mathcal{O} \qty( \frac{\log^2 d}{d^2} ) }. \label{rpplad}
\ea
For small extremal black holes $r_h \ll L$,  this is already a well-defined large-$d$ expansion.  It is worth noting that for small near-extremal black holes in AdS do not suffer from a negative specific heat unlike their uncharged counterparts. For large AdS black holes with $r_h \gg L$,  we need the additional condition for the validity of the expansion \eqref{rpplad} that $d \gg \log (r_h^2 /L^2)$.

Note that the location of the photon sphere defined by \eqref{rpdef} does not depend on the angular momentum parameter $\ell$ and can exist for any value of $\ell$.  The value of the energy $E$ is given by substituting $r = r_{p+}$ in eq.  \eqref{radeq} and setting $\dv*{r}{\la} = 0$.  In a large $d$ expansion,  we find
\baa
E^2 = \ell^2\pqty{ \frac{1}{L^2} + \frac{1}{r_h^2}  } - \frac{2 \ell^2}{r_h^2} \frac{1 + \log  d + \log(1 + \frac{r_h^2}{L^2}) }{d} + \mathcal{O} \qty( \frac{\log^2 d}{d^2} ).  \label{eonps}
\ea

The foregoing results also make it clear why there is no photon ``plane'' outside the horizon for a planar black brane.  The planar brane in a fixed $d$ can be obtained by taking the limit $r_h / L \to \infty$.  This operation,  eq.  \eqref{rphpl} tells us,  sends the unstable photon sphere to asymptotic infinity.  In fact,  starting with the exact form of the metric we find that only the branch \eqref{innerphots} is present.  The potential is a monotonic increasing function of $r$ outside the horizon.

We have so far discussed an extremal black hole exactly at zero temperature. The exact zero-temperature case can sometimes lead to unwelcome pathologies, even at the classical level. With this in mind,  we now consider a small temperature obtained by adding a small amount of mass above extremality,  $M = M_{\rm ext} + \delta M$, but keeping the charge fixed at $Q = Q_{\rm ext}$, which corresponds to working in the canonical ensemble \cite{Moitra:2019bub, Moitra:2022ooc}.  Addition of the mass makes the horizon non-degenerate and the horizon is shifted to $r = r_h + \delta$,  where the shift of the horizon is given by
\baa
\delta M = \frac{r_h^{d-2}}{2 L_2^2} \delta^2 \pqty{ 1 - (\alpha - d +2) \frac{\delta}{r_h} + \cdots    },   \label{delmd}
\ea 
where 
\baa
\alpha =  \frac{1}{3} \frac{(d-1) [ 3(d-2)^2 L^2 + d(3d-5) r_h^2  }{(d-2)^2 L^2 + d(d-1) r_h^2 } \label{alpha}
\ea
remains finite in $d \to \infty$ and $r_h \to \infty$ or $L \to \infty$ limits. The near-horizon metric reads
\baa
\dd{s}^2 = - \frac{(r-r_h)^2  - \delta^2}{L_2^2} + \frac{L_2^2}{(r-r_h)^2 - \delta^2} \dd{r}^2 + r_h^2 \dd{\Omega}_{d-1}^2.  \label{nhmet}
\ea
For a fixed $d$,  a black hole is defined to be nearly extremal if $\delta / r_h \ll 1$. As we take the large $d$ limit, however,  we note from \eqref{delmd} that we need to impose a stronger constraint for the sub-leading corrections to be small,
\baa
\frac{d \delta}{r_h} \ll 1.  \label{newlim}
\ea
How does this translate into a condition for the temperature?  The temperature can be read off from the metric \eqref{nhmet} to be
\baa
T = \frac{\delta}{2\pi L_2^2} \approx \frac{d^2 \delta}{2\pi} \pqty{ \frac{1}{L^2} + \frac{1}{r_h^2}  }.  \label{tempdef}
\ea
On the other hand,  the scale of the near-horizon symmetry breaking \cite{Moitra:2019bub} is given by
\baa
\mathcal{J} =  \frac{r_h}{L_2^2}  \approx d^2 r_h  \pqty{ \frac{1}{L^2} + \frac{1}{r_h^2}  }.  \label{jdef}
\ea
Thus $T \ll \mathcal{J}$ is always satisfied and we are firmly in the low-temperature regime.    

The discussion so far has been classical.   There is, however,  the possibility that at large $d$, the temperature is so small that quantum corrections to the near-extremal description become important \cite{Preskill:1991tb}.   Condition that this does not happen is given by
\baa
T \gg \frac{\mathcal{J}}{S_0}, \label{semc}
\ea
where $S_0$ is the ground-state entropy of the black hole.  In terms of the $d+1$-dimensional Planck length $L_{\rm Planck}$,  we can write this entropy as
\baa
S_0 \sim \frac{\pi^{d/2}}{\Gamma(d/2)} \qty( \frac{r_h}{L_{\rm Planck} } )^{d-1} \approx \qty( \frac{ 2 \pi e }{d} )^{d/2} \qty( \frac{r_h}{L_{\rm Planck} } )^{d-1}.  \label{gsent}
\ea
For $ d \gg 2\pi (r_h / L_{\rm Planck})^2$, this entropy becomes exponentially small in $d$.  Thus,  to avoid the possibility of quantum corrections, we could either work with a $d$ close to this value or choose a scaling of the $d$-dimensional Planck length to be $L_{\rm Planck} \sim 1/\sqrt{d}$ so that $S_0$ becomes an exponentially large number. In such a scenario,  \eqref{semc} becomes trivial and we can effectively take the zero-temperature limit. In any case,  as will be discussed in section \ref{sec-disc},  even in the presence of quantum corrections at $T \sim \mathcal{J}/S_0$,  the essential features of the geometry are expected to remain unchanged.

Let us now look at the photon ring under such a finite-temperature deformation. Assuming the limit \eqref{newlim},  we find corrections to the location of the photon ring,
\baa
r_{p +} (M_{\rm ext} + \delta M ,  Q_{\rm ext}) = r_h \bqty{ 1 + \frac{\log d}{d}  + \frac{\log(1 + \frac{r_h^2}{L^2})}{d}  +  \frac{d^2 \delta^2}{r_h^2} + \cdots }. \label{rpplad2}
\ea
Thus, the photon ring is shifted only slightly relative to the horizon,  as one would expect.

\subsection{Massless Geodesics with Charge}\label{ss-charged}

We now consider massless geodesics which are also charged under the same $\mathrm{U} (1)$ as the black hole. with a charge $q$.  There are several motivations for studying the motion of such particles.  First of all,  as we will show below,  such a geodesic is an excellent approximation for describing a charged field with any mass in the geometric optics limit, with the charge scaling in the same way as the eikonal.  Secondly,  under a Kaluza-Klein reduction,  a rotation in a compactified direction corresponds to a charge in the lower-dimensional effective theory.  The ordinary null geodesic equation for a rotating black hole in higher dimension would become a charged null equation in lower dimensions.   

In the present context,  consideration of such a geodesic leads to a surprising outcome --- one can have circular null geodesics close to the horizon in any dimension.   This geodesic motion of the particle is governed by the equation
\baa
\qty( \dv{r}{\lambda} )^2 + V(r) = 0, \label{d2r1}
\ea
with
\baa
V(r) \equiv \ell^2 \frac{f(r)}{r^2} - \qty[E + q A_t (r) ]^2. \label{vrdef}
\ea
In the limit $q \to 0$,  we recover eq.  \eqref{radeq}.  For a non-zero $q$, however,  the radial potential \eqref{vrdef} is quite a different and fantastic beast.  For one,  a non-zero charge $q$ explicitly selects a particular affine parametrization.  In the previous case,  $\ell$ and $E$ varied simultaneously under a rescaling of the affine parameter $\lambda$.  This rescaling symmetry is broken by $q$.  The affine parameter in this case can be regarded as a certain massless limit for the proper time corresponding to a massive particle.   This potential has other interesting features.  In the solution for a circular orbit in \eqref{radeq} the angular momentum $\ell$ can be taken to be completely arbitrary --- eq. \eqref{rpdef} determined the photon orbit radius completely in terms of black hole parameters.  In this case,  if the photon orbit is at $r =  r_p$,  the circular orbit equations $V (r_p ) = 0 = V'(r_p)$ lead to the solution
\baa
E &=   -  \frac{2q f(r_p) A'_t (r_p) }{2f(r_p) -r_p f'(r_p) } - q A_t (r_p), \\
\ell^2 &= \frac{4 q^2 r_p^4 A'_t (r_p)^2 f(r_p)  }{(2f(r_p) -r_p f'(r_p) )^2}.  \label{el2ch} 
\ea
Thus,  $r_p$ is a parameter of our choice and we can have a charged massless orbit anywhere we like (note that $\ell^2$ is positive outside the horizon).  

It is worth pointing out that the quantity $2f(r_p) -r_p f'(r_p)$ appears in the denominator of both the formulas in \eqref{el2ch}.   These formulas thus fail to work at the value of $r_p$ at which this quantity vanishes.   As the astute reader would have noticed,  this happens precisely at the location of the neutral photon ring,  defined by \eqref{rpdef}.  Inside (outside) the neutral photon ring,  the quantity $2f(r_p) -r_p f'(r_p)$ is negative (positive).  Thus,  close to the horizon of an extremal black hole $r_p \gtrapprox r_h$ this quantity is negative,
\baa
2f(r_p) -r_p f'(r_p) = -  \frac{2 r_h (r_p - r_h)}{L_2^2} + \mathcal{O} \pqty{ (r_p - r_h)^2 }. \label{frp2}
\ea
Since $E + q A_t (r) = f(r) \dv*{t}{\lambda}$,   for future-directed paths $\dv*{t}{\lambda} > 0$,  we must have $q Q > 0$ --- i.e.,  the charges must be of the same sign.   On the other hand,  outside of the neutral photon ring one must have $q Q < 0$ for the geodesic to be future-directed.

The second derivative of the potential \eqref{vrdef} tells us about the stability of the orbit and we find,
\baa
V''(r_p) = - \frac{2 (d-1) \left(d r_h^2+(d-2) L^2\right) \left(2 (d-1) d r_h^2+3 (d-2)^2 L^2\right)}{3 L^2 r_h^2 \left((d-1) d r_h^2+(d-2)^2 L^2\right)} q^2 \frac{r_p -  r_h}{r_h} + \cdots ,\label{vdprp}
\ea
where the ellipsis denotes higher order terms in $(r_p - r_h)$.
Thus,  the second derivative is necessarily negative right outside the horizon, signifying that it is an unstable photon orbit.  It turns out that,  except for the divergent behavior on the neutral light ring,  in (near-)extremal geometries,  $V''(r_p)$ is negative everywhere outside the horizon for $d \geq 4$.   An interesting phenomenon occurs  for $ d=3$: for finite values of $L$,  $V''(r_p)$ becomes positive for sufficiently large values of $r_p$,  which signifies the existence of stable charged photon orbits of large radius.  However,  this peculiar behavior does not survive in the two extremes of asymptotically flat or planar limits and in these cases we get $V''(r_p) < 0$ everywhere.

Let us also write down the energy and angular momentum \eqref{el2ch} for orbits close to the horizon.   Taking a positive $Q_{\mathrm{ext}}$, we find,
\baa
E &\approx \frac{q}{3} \sqrt{\frac{(d-1)(d-2)}{2}}  \pqty{ \frac{1}{r_h^2} + \frac{d}{d-2} \frac{1}{L^2} }^{1/2} \frac{ 3(d-2)^2 L^2 + 2 d (d-1) r_h^2  }{(d-2)^2 L^2 + d(d-1) r_h^2 } (r_p - r_h)^2, \\
\ell^2 &\approx  q^2 r_h^2 \frac{d-1}{2} \frac{(d-2) L^2 + d r_h^2 }{(d-2)^2 L^2 + d(d-1) r_h^2}.  \label{elexp}
\ea

It is worth emphasizing that unlike their uncharged counterparts,  these orbits exist even for planar black branes.  In this aspect,  these are different from the innermost stable circular orbits (ISCO) of massive charged particles in AdS,  which do not exist for the planar geometry,  as noted in \cite{Moitra:2023yyc} (see also \cite{Berenstein:2020vlp}).

Before we end this section,  let us see how this potential can be derived in the geometric optics approximation.  Let us consider the Klein-Gordon equation for a charged field with mass $m$,
\baa
D_\mu D^\mu \Psi - m^2 \Psi = 0,
\label{kgch}
\ea
where $D_\mu \equiv \nabla_\mu - \mi q A_\mu$ is the gauge-covariant derivative.
Let us consider wavelike solutions 
\baa
\Psi = \mathcal{A} e^{\mi \mathcal{S} / \hbar},  \label{waveik}
\ea
where $\hbar$ is a formal small parameter and both the amplitude $\mathcal{A}$ and the eikonal $\mathcal{S}$ are assumed to admit a nice power series expansion in $\hbar$.  Let us consider that the charge goes as $q/\hbar$ in this parameter. Noting that the gauge-covariant derivative then becomes $D_\mu = \nabla_\mu - \mi q A_\mu / \hbar$,  we obtain
\baa
- \frac{1}{\hbar^2} (\nabla_\mu \mathcal{S}  - q A_\mu )^2 + \frac{\mi}{\hbar} \pqty{\frac{2 \nabla^\mu \mathcal{A}}{\mathcal{A} } + \nabla^\mu } ( \nabla_\mu \mathcal{S}  - q A_\mu)  + \pqty{ \frac{\nabla^2 \mathcal{A}}{\mathcal{A}} - m^2} = 0. \label{eik2}
\ea
The leading order equation in $\hbar$,
\baa
(\nabla_\mu \mathcal{S} - q A_\mu )^2 = 0 \label{hamilj}
\ea
is precisely the form of Hamilton-Jacobi equation for a massless charged particle moving in an electromagnetic field.  Defining $p_\mu \equiv \nabla_\mu \mathcal{S}$ and $k_\mu \equiv p_\mu - q A_\mu$, we see that $k^\mu$ is a null vector $(k_\mu k^\mu = 0)$ that satisfies the charged geodesic equation,
\baa
k^\mu \nabla_\mu k_\nu = -  q k^\mu F_{\mu \nu}.  \label{geod}
\ea
The null geodesic equation \eqref{d2r1} would thus emerge universally under such a limit.

\section{Null Particle Trajectories and Quasi-Normal Modes}\label{sec-qnm}

Let us consider the Klein-Gordon equation \eqref{kgch} for a massive charged scalar field in a background of the form \eqref{metric} with the gauge field \eqref{gauge}.  We can write the mode solutions as
\baa
\Psi (x^\mu)  = e^{ - \mi \omega t } Y_{\ell \cdots } (\Omega) \frac{\psi(r)}{r^{\frac{d-1}{2}}},  \label{psi}
\ea
where $Y_{\ell \cdots } (\Omega)$ denotes the $(d-1)$-dimensional spherical harmonics. 
This leads to an wave-mechanical equation expressed in terms of the coordinate $z$ \eqref{regwhee},
\baa
-\dv[2]{\psi}{z} + V(z) = 0,  \label{waveeq}
\ea
where the potential is given by
\baa
V = f(r) \pqty{ \frac{\ell(\ell + d -2)}{r^2} + m^2 + \frac{(d-1)(d-3)}{4} \frac{f(r)}{r^2} + \frac{d-1}{2r} f'(r)  } - \pqty{ \omega + q A_t  }^2. \label{fieff2}
\ea
In the AdS/CFT correspondence,  a bulk field of mass $m$ is represented as a boundary operator of scaling dimension $\Delta_+ = d/2 + \nu$ (in the standard quantization which we consider throughout this article),  where $\nu$ is given by
\baa
\nu^2 = m^2 L^2  + \frac{d^2}{4}.  \label{nudef}
\ea
As in \cite{Festuccia:2008zx, Dodelson:2023nnr}, we will find it convenient to use $\nu$ instead of $m$. We can see how the potential \eqref{vrdef} arises from this equation by taking large $\ell$,  $\omega$ and $q$,  with all of them scaling in a similar manner.  In the $z$-coordinate,  the asymptotic fall-off of the scalar is given by
\baa
\psi \sim \mathcal{A} z^{ \frac{1}{2} - \nu } + \mathcal{B} z^{ \frac{1}{2} + \nu } + \cdots,
\ea
where $\mathcal{A}$ and $\mathcal{B}$ are associated with the non-normalizable and normalizable modes respectively.  We can find the retarded Green function from the ratio $\mathcal{B} / \mathcal{A}$.  The quasi-normal poles can be read off from the zeros of $\mathcal{A}$.

\begin{figure}
\centering
\begin{subfigure}{.45\textwidth}
  \centering
  \includegraphics[width=0.95\linewidth]{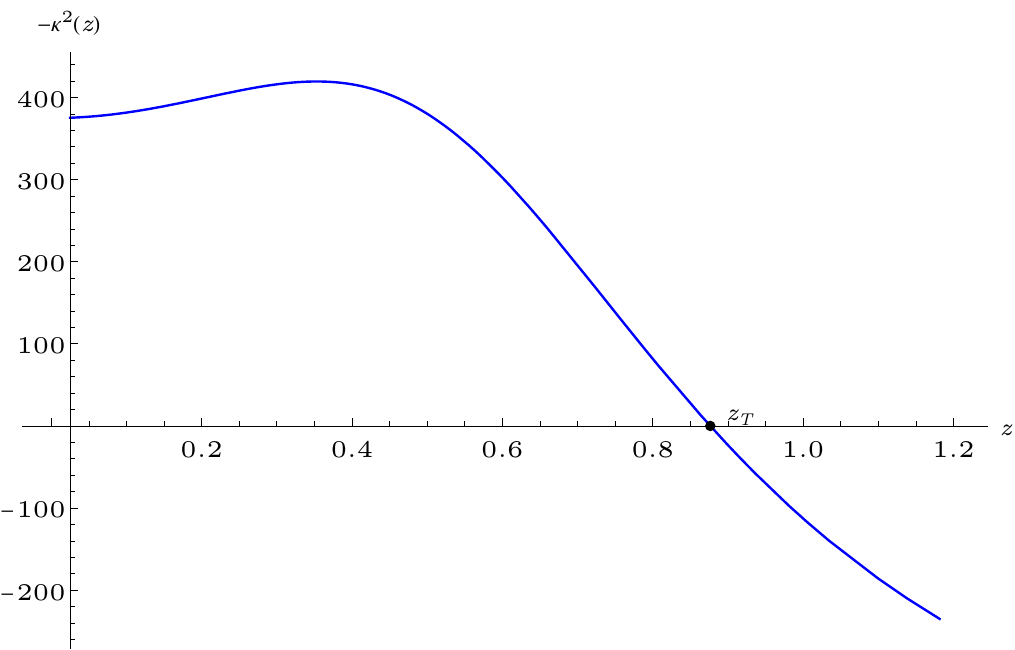}
\caption{$\qh = - 20$,  $\kh = 20$, $\oh = 25$}
\label{fig-turning1}
\end{subfigure}%
\begin{subfigure}{.45\textwidth}
  \centering
  \includegraphics[width=0.95\linewidth]{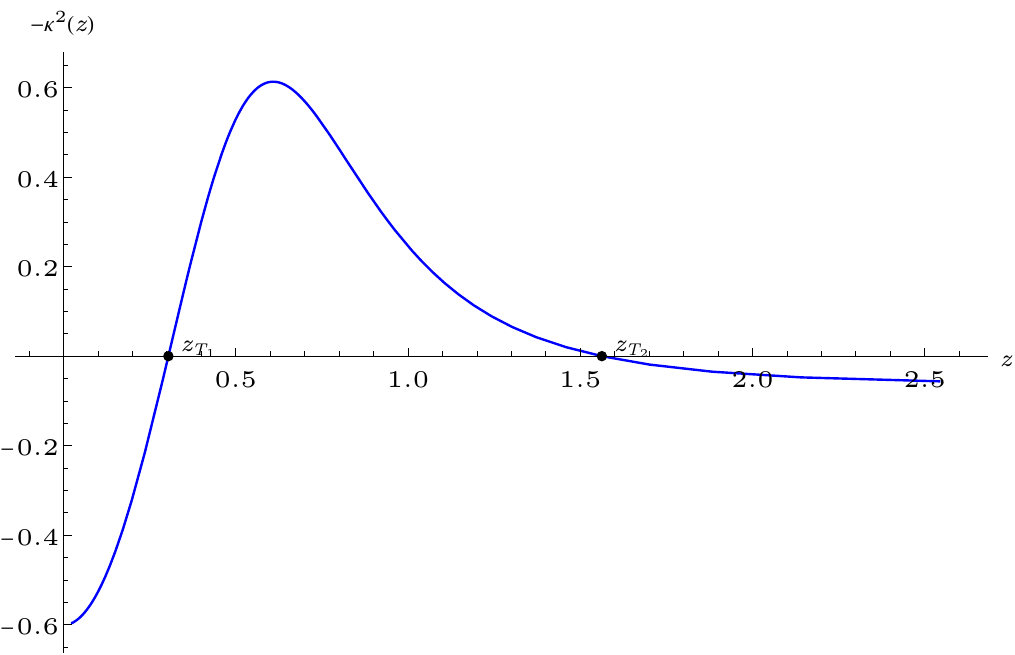}
  \caption{$\qh = 1$,  $\kh = \sqrt{7/3}$, $\oh = 23/108$}
\label{fig-turning2}
\end{subfigure}
\caption{Plots of the eikonal potential $-\kappa^2 (z)$ with one or two classical turning points in extremal geometries with parameters $d=4$,   $r_h  = 1 = L$. The particle can be thought to move in these potentials at zero energy $\kappa^2 = 0$.}
\label{fig-turning}
\end{figure}

\subsection{Null Particles in Geometries with One Real Turning Point}\label{ss-onetp}

For null trajectories in black hole geometries,  depending on the parameters of the trajectory,  there can be one or more turning points,  as illustrated with specific examples in Figure \ref{fig-turning}. We will first consider the case where the null trajectory has only one turning point.  This will not only serve as a simple illustration of the methods involved, it will also allow us to make comments about interesting physical situations such as CFTs on $\mathbb{R}^{1,d-1}$,  which we discuss next.

\subsubsection{Planar Black Brane Geometry}\label{ss-planar}

The red-shift factor in this case is given by simply dropping the ``1'' in eq.  \eqref{redshift}
\baa
f(r) = \frac{r^2}{L^2} - \frac{2M}{r^{d-2} } + \frac{Q^2}{r^{2d-4}}.  \label{bfpla}
\ea
The metric has the form, instead of \eqref{metric},
\baa
\dd{s}^2 = - f(r) \dd{t}^2 + \frac{\dd{r}}{f(r)} + r^2 \dd{\vec{x}}_{d-1}^2 \label{metric2}.
\ea
The spherical harmonics would now be replaced by standard Fourier modes of (dimensionless) momentum $k$,   $Y_{\ell \cdots} (\Omega) \to \exp( \mi k x_1)$.  Using rotational invariance,  we have oriented the momentum in a particular spatial direction $x_1$.  The eigenvalue $-\ell (\ell + d-2)$ would now be replaced by $-k^2$.  We would thus have the potential,
\baa
V = f(r) \pqty{ \frac{k^2}{r^2} + \frac{\nu^2 - \frac{1}{4}}{L^2} + \frac{(d-1)^2 M}{2r^{d}} - \frac{(3d-5)(d-1) Q^2}{4r^{2d-2}}  } - \pqty{ \omega + q A_t   }^2.  \label{potpla}
\ea
We now take the eikonal limit of the wave equation.  We first define the parameters
\baa
k = \La \kh, \quad \om = \La \oh,  \quad q = \La \qh  \label{lamdef}
\ea
and take the limit $\La \gg 1$ (unhatted quantities are related to hatted quantities in the same way in the remainder of this article),  where $\La$ is related to the inverse of the small parameter $\hbar$.  We write the wave equation as
\baa
- \dv[2]{\psi}{z} -  \Lambda^2 \kappa^2 (z) =0,  \label{waveq1}
\ea
which is suitable for a JWKB approximation.   The function $\kappa^2 (z)$,  whose negative we shall refer to as the \emph{eikonal potential} in the remainder of the article,  is given by
\baa
\kappa^2 (z) = \pqty{ \oh + \qh A_t  }^2  -  f(r) \frac{\kh^2}{r^2} .  \label{kap1}
\ea
The classically allowed (forbidden) region corresponds to $\kappa^2 (z) > 0$ ($<0$). The classical turning points are located at the value  of $z$ for which $\kappa(z) = 0$.

From the form of the potential above,  it is obvious that near the black hole horizon $z \to \infty$,  the potential is of the form
\baa
\kappa^2 (z) \approx  \oh^2.  \label{kapho}
\ea
since both $f(r)$ and $A_t (r)$ vanish on the horizon.  Near the asymptotic boundary $r \to \infty$,  the eikonal approximation breaks down and we have to retain the effect of the mass.  The potential in this case is given by 
\baa
V (z \to 0) \approx  \frac{\nu^2 - \frac{1}{4}}{z^2} + \Lambda^2 \rho^2, \label{potasy1}
\ea
where,
\baa
\rho^2 \equiv \frac{\kh^2}{L^2}  - (\oh + \qh \mu)^2 \label{defrho}.
\ea
For sufficiently small $z$,  the positive first term always dominates over the second irrespective of the sign of $\rho^2$.  The conformal boundary is thus always in the classically forbidden region.

One might be concerned that since we are working with a large value of $q$,  it might lead to a violation of the Breitenlohner-Freedman bound \cite{Breitenlohner:1982jf} for the near-horizon $\mathrm{AdS}_2$ region, thus triggering an instability.   For the planar brane, we find that the $\mathrm{AdS}_2$ scaling dimension satisfies
\baa
\Delta^{(2)} (\Delta^{(2)} - 1) = L_2^2 \pqty{ m^2 - \frac{1}{2} q^2 + \frac{k^2}{r_h^2}  }. \label{bfbound}
\ea
Even if there is an instability for large values of $q$, that is a different question from what we are interested in right now.  Furthermore,  modes with sufficiently large values of $k$ do not trigger the instability anyway.

We will now consider situations where the JWKB eikonal potential $\kappa^2 (z)$ will have exactly one real turning point in $z$,  say at $z= z_T$ (or $r= r_T$).   The JWKB wave function with ingoing boundary condition on the horizon reads
\baa
\psi (z) = \frac{1}{\sqrt{\Lambda \kappa (z) }} \exp( \mi \La \int_{z_T}^{z} \dd{z}\kappa (z)    ),  \qquad (z > z_T) \label{JWKBpal}
\ea
where $\kappa (z)$ is the positive root of eq.  \eqref{kap1}. 

Black hole quasi-normal modes using complex JWKB methods, in which one complexifies the radial coordinate,  have been studied extensively in the literature \cite{Festuccia:2008zx}.
As mentioned in the introduction,  in this work, we propose a relatively simple $\mi \e$-prescription which,  when combined with appropriate asymptotic expansions \cite{Dingle:1973} of relevant functions (such as  modified Bessel or Airy functions),  reproduces the correct results in the literature.  This $\mi \e$-prescriptions involves an analytic continuation of the large JWKB parameter $\La$ which ensures that the JWKB solution \eqref{JWKBpal} is well-behaved on the horizon 
\baa
\Lambda \to \Lambda ( 1 + \mi \e), \qquad \e >0. \label{iep1}
\ea
The JWKB solution in the classically forbidden region is then given by
\baa
\psi (z) = \frac{1}{\sqrt{\Lambda | \kappa (z) |} } \bqty{ A \exp(\La \int_z^{z_T} \dd{z} |\kappa(z) |   )+ B \exp(- \La \int_z^{z_T} \dd{z} |\kappa(z) |   )  }. \label{JWKBpd}
\ea
One must exercise caution in superposing growing and  decaying exponentials in the classically forbidden region.  It is however the competition of these exponentials that is responsible for determining the quasi-normal poles of the retarded Green function,  as emphasized in  \cite{Festuccia:2008zx}.

A careful analysis with asymptotic forms \cite{Dingle:1973}  of Airy functions gives the connection formula, $A =  \exp(-\mi \pi /4) = -\mi B$.  We obtain the asymptotic solution in terms of modified Bessel functions of first $(I_\nu)$ and second $(K_\nu)$ kinds,
\baa
e^{ \frac{\mi \pi}{4} } \psi (z) = \sqrt{2\pi} \mi e^{- \theta}  \sqrt{z} I_\nu (\Lambda \rho z) + \sqrt{\frac{2}{\pi}} (e^\theta + e^{\mi \pi \nu} e^{-\theta} ) \sqrt{z}   K_\nu (\Lambda \rho z),  \label{besol}
\ea
where $\te$ refers to the integral,
\baa
\te \equiv \La \int_0^{z_T} \dd{z} |\kappa (z)|. \label{tedef}
\ea
Setting the non-normalizable component of \eqref{besol} to zero gives us the quantization condition,
\baa
2 \te = \mi \pi ( 2n +1  + \nu), \quad n \in \mathbb{Z}_{\geq 0} \label{quancon}.
\ea
This is precisely the quantization condition obtained in \cite{Festuccia:2008zx} by a detailed analysis.  The fact that $n \geq 0$ has to do with the fact that these modes correspond to a Stokes half-line. This lends further support to the $\mi \e$-prescription that we have used. We can obtain the retarded Green function from \eqref{besol} by evaluating the normalizable and non-normalizable modes.  We find,
\baa
G_R (\omega,  k)  = \frac{\Gamma(-\nu)}{\Gamma(\nu)}  \pqty{ \frac{\Lambda^2 \rho^2}{4} }^\nu e^{-\mi \pi \nu} \frac{\cosh( \theta + \frac{\mi \pi \nu}{2}  )}{\cosh( \theta - \frac{\mi \pi \nu}{2}  )}. \label{greta}
\ea
This Green function is closely related to the one obtained by the authors of \cite{Dodelson:2023nnr} at infinite volume with complex spin.  The authors  had noted that the QNM obtained from imaginary spin agrees with the results by a na\"ive analytic continuation from real spin (which we will soon compute).  At imaginary spin,  the intermediate steps are superficially different.  Our $\mi \e$-prescription makes it clear why this had to happen: we should deform $\Lambda$ in the upper half plane.  In fact, by sending $\Lambda \to \Lambda \mi$ in the $\cosh$ functions in \eqref{greta},  we recover exactly the same expression for the retarded Green function as in \cite{Dodelson:2023nnr}.

To calculate the QNM frequencies in the JWKB approximation,  we should explicitly evaluate the integral in \eqref{tedef},
\baa
\theta = \La \int_{r_T}^\infty \frac{\dd{r}}{f(r)} \sqrt{f(r) \frac{\kh^2}{r^2} - \pqty{ \oh + \qh \mu - \qh c_d \frac{Q}{r^{d-2}} }^2 }, \label{thein}
\ea
where, as mentioned previously,  $r_T$ is the turning point,  at which the argument of the square-root vanishes linearly.  We can express $\oh + \qh \mu$ in terms of $r_T$, 
\baa
\oh + \qh \mu = \qh c_d \frac{Q}{r_T^{d-2}} + \frac{\kh}{r_T} \sqrt{f(r_T)}.  \label{ort}
\ea
Unfortunately,  the integral in \eqref{thein} does not allow a closed-form solution,  unlike in the corresponding scenario in \cite{Dodelson:2023nnr}.  For large values of the turning point,  however,  we can use a perturbative approach, which turns out to be more instructive.  For large $r_T$,  we can write from \eqref{ort},
\baa
r_T \approx  \bqty{ \frac{\frac{k}{L} - (\om+ q \mu)}{ -q c_d Q }  }^{-\frac{1}{d-2}}.  \label{rtapprox}
\ea
Note that this immediately makes clear that $\qh Q < 0$ for this solution to make sense.  It is not possible for the eikonal potential in \eqref{kap1} to change sign for a real value of  $r$ if $q$ and $Q$ do not have opposite signs.

To evaluate the integral,  we transform to a new variable,
\baa
y = \frac{r_T}{r}.  \label{ydef}
\ea
In terms of this new variable,
\baa
\theta = \Lambda r_T \int_0^1 \frac{\dd{y}}{y^2 f(r_T / y)} \sqrt{ f(r_T/ y) \frac{\kh^2 y^2}{r_T^2}  - \pqty{ \frac{\kh}{r_T} \sqrt{f (r_T)} + \frac{\qh c_d Q }{ r_T^{d-2}   } \pqty{ 1 - y^{d-2} } }^2  }. \label{tedeva}
\ea
We will now do a large-$r_T$ expansion of the integrand.  In, doing so, we find the leading term to be
\baa
\theta &= \frac{\Lambda L^2}{r_T} \sqrt{-  \frac{2 \kh \qh c_d Q}{L r_T^{d-2}}} \int_0^1 \dd{y} \sqrt{1 -  y^{d-2}} \\ &= \frac{L^{3/2}}{d} B \pqty{ \frac{1}{2} ,  \frac{1}{d-2}  } \sqrt{- \frac{2k q c_d Q }{r_T^{d}}}    \label{tedeva2},
\ea
where $B (x, y)$ is the Euler beta function.  Using the asymptotic expression \eqref{rtapprox} for $r_T$ and the quantization condition \eqref{quancon},  we obtain the quasi-normal frequencies,
\baa
(\om + q \mu)L = k -  \bqty{  \frac{\mi \sqrt{\pi}}{ 2 \sqrt{2} k^{1/2} } ( 2n + 1 + \nu)  \frac{d \Gamma( \frac12 + \frac{1}{d-2}  )}{ \Gamma(\frac{1}{d-2} )} }^{\frac{2(d-2)}{d} }  \qty( - q L c_d \frac{Q}{L^{d-2}})^{\frac{2}{d} }. \label{qnm1}
\ea
The fact that $\omega + q \mu$ appears in the expression instead of simply $\omega$ is on account of the particular gauge we have chosen to work in.  One can interpret $\omega + q \mu$ as the boundary energy of the charged operator. The $k$-dependence of the correction term for the charged operator is $k^{ - \frac{d-2}{d} }$ which is smaller than the scaling observed for uncharged black holes \cite{Festuccia:2008zx, Dodelson:2023nnr},  going like  $k^{- \frac{d-2}{d+2} }$.   This implies that for such charged operators,  the relaxation time-scale is longer and it takes a parametrically longer time to reach equilibrium for charged operators.  At a qualitative level, the difference arises because in the eikonal limit of the kind considered here,  the charge of the bulk particle is not suppressed by the red-shift factor like the mass is.  As noted before,  this formula is true for $qQ <0$. When we translate this as a statement for rotating black holes and substitute an appropriate value of $q$,  this formula work for \emph{counter-}rotating modes.  It is also worth noting that in $d=3$ these quasi-normal modes make an obtuse angle with positive real $\omega$-axis. This will obviously not continue indefinitely and the expansion will eventually break down.

\begin{figure}
\centering
\includegraphics[width=\linewidth]{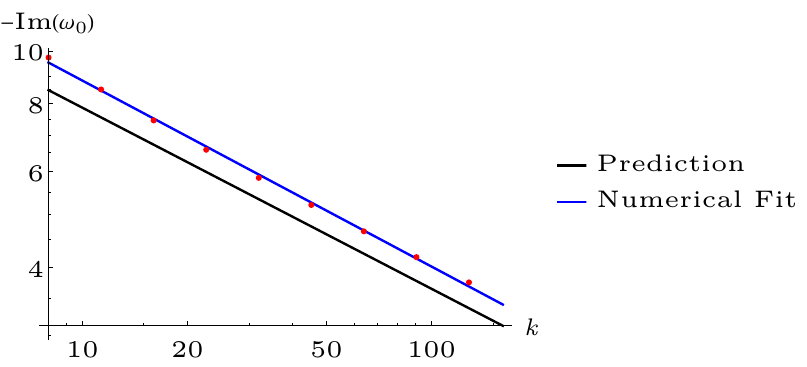}
\caption{The charged quasi-normal modes are plotted as a function of the momentum,  for $d=3$, $q = -12$, $\nu =3/2$ and $r_h = 1 = L$. Red dots denote specific numerical results.}
\label{fig-charged}
\end{figure}

The new scaling is well-supported by a numerical evaluation of the quasi-normal spectrum,  as shown in Figure \ref{fig-charged} for an extremal black hole.  
{ It is worth explaining the result in some detail --- at first sight,  it appears that there is a  substantial disagreement between the prediction and the curve obtained by fitting the data.  
The formula \eqref{qnm1},  obtained from the JWKB approximation,  is best suited for describing the semiclassical regime,  i.e.,  high overtone numbers $n\gg 1$.   In contrast,  using the package \cite{Jansen:2017oag},  we were able to reliably evaluate the QNMs in Figure \ref{fig-charged} for the \emph{fundamental} mode, $n=0$,  for which we do not expect a good agreement anyway.   Our interest is to find evidence for the new $k$-scaling,  keeping everything else fixed.  In spite of the lack of an exact agreement in the values of frequencies,  we see that the observed and predicted curves share the same \emph{slope} on the log-log plot.  It is therefore quite striking that the predicted $k$-scaling is borne out by the fundamental mode.  We expect the two curves in Figure \ref{fig-charged} to coincide for high enough values of the overtone number $n$ (with a correspondingly higher value of $k$ for the perturbative expansion in \eqref{qnm1} to be valid). }

Note further that we have not assumed a particular value of $Q$ in this derivation.  If the black hole is extremal,  we can substitute $Q = \sqrt{\frac{d}{d-2}} r_h^{d-1}/L$ to find the extremal limit.   Let us also consider the extremal limit together with $d \gg 1$,  in this case,
\baa
(\om  + q \mu)L  \approx  k + \frac{\pi^2 \mu^2}{2 k} (2n +1 + \nu)^2 ( - q L )^{ \frac{2}{d} } \pqty{ 1  - \mi \frac{2\pi}{d} + \cdots} ,  \label{qnmldq}
\ea
where we have shown the first non-vanishing imaginary part and the dots denote higher order corrections.
As expected,  it is the chemical potential that sets the scale of the QNMs.  The imaginary parts,  not shown here,  would be suppressed in $d$.  In taking the large-$d$ limit,  we have to be careful that the approximation in obtaining the eikonal potential \eqref{kap1} does not break down, in particular $k^2$ and $q^2$ should grow faster than $d^2$ for this approximation to work well.   It would be interesting to explore this regime numerically.

One can also wonder about the analog of the formula \eqref{qnm1} for $qQ > 0$.  One obvious problem, as noted before,  is that there is real turning points located arbitrarily close to the conformal boundary.  One can, however,  consider \emph{complex} turning points and do a complex rotation in $q$.   A na\"ive analysis suggests that the quasi-normal modes are all real,  with no imaginary component --- i.e.,  they are normal modes.  It would be interesting to explore this regime from a numerical perspective.

\subsubsection{The Zero Probe Charge Limit}\label{ss-zeroq}

\begin{figure}
\centering
\begin{subfigure}{.5\textwidth}
  \centering
  \includegraphics[width=0.95\linewidth]{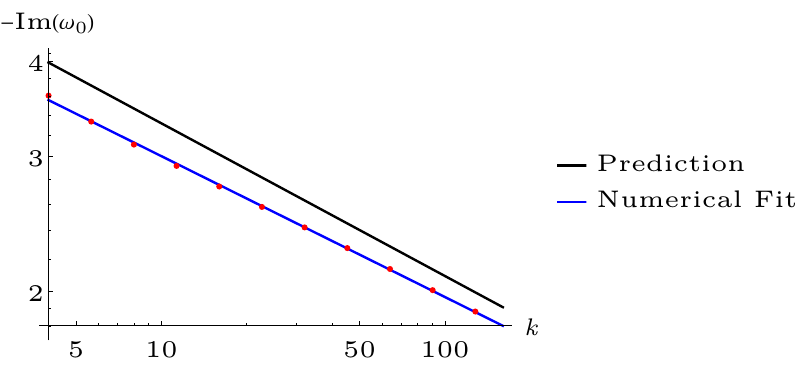}
  \caption{$k$-dependence}
\label{fig-neutralqnm1}
\end{subfigure}%
\begin{subfigure}{.5\textwidth}
  \centering
  \includegraphics[width=0.95\linewidth]{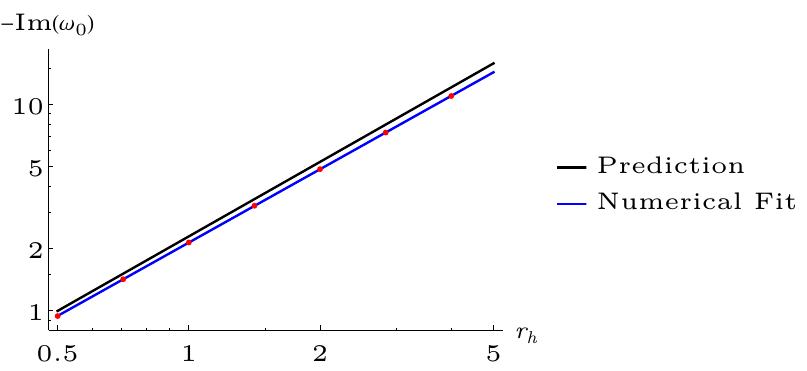}
  \caption{$\mu$-dependence}
\label{fig-neutralqnm2}
\end{subfigure}
\caption{The neutral quasi-normal modes are plotted as a function of the momentum,  for $d=3$,  $\nu = 3/2$ and $r_h = 1 = L$.  Red dots denote specific numerical results.}
\label{fig-neutralqnm}
\end{figure}

It is worth discussing the $q \to 0$ limit in some detail and pointing out the relation with existing literature. The na\"ive $q\to 0$ limit of the previous section does not work, because the perturbation theory set-up for large-$r_T$ breaks down in this limit.  Equation \eqref{ort} with $q \to 0$ still remains valid,  but instead of \eqref{rtapprox} yields the following solution for the turning point at large values of the radial coordinate,
\baa
r_T \approx \bqty{ \frac{ \frac{k}{L} -  \om}{k L M } }^{- \frac{1}{d} }  \label{rtapprox2}.
\ea
Evaluating \eqref{tedeva} with $\qh = 0$ in the large-$r_T$ limit yields
\baa
\te =  \frac{k L^2}{d+2} B \pqty{ \frac{1}{2},  \frac{1}{d} } \sqrt{ \frac{2M}{r_T^{d+2}} }. \label{theta}
\ea
Using the approximation \eqref{rtapprox2} and the quantization condition \eqref{quancon},  we obtain the quasi-normal frequencies
\baa
\omega L = k-  k^{- \frac{d-2}{d+2}} \pqty{ \mi \sqrt{\frac{\pi}{2}} \frac{\Gamma(\frac32 + \frac1d)}{\Gamma( 1+ \frac1d)}  (2n +1 + \nu) }^{\frac{2d}{d+2}} \pqty{\frac{M}{L^{d-2}}}^{\frac{2}{d+2}} .\label{disp2}
\ea
When we take the limit,  $Q \to 0$,  this agrees with the results in \cite{Festuccia:2008zx} and \cite{Dodelson:2023nnr}, after a suitable analytic continuation.  To this order,  our result is true even  for a non-zero $Q$,  provided $r_T$ is sufficiently large.  In fact, our work shows that the answer, up to this order, depends only on the mass.  In the absence of any black hole charge,  the mass parameter depends only on the temperature $M \sim  T^d$ and thus the second term above goes like $T^\frac{2d}{d+2}$,  as found in \cite{Festuccia:2008zx}.  In case no charge was present,  the zero-temperature limit would have implied zero mass and hence an empty AdS spacetime with only the first term of \eqref{disp2}.  In the presence of global charges on the CFT,  this relation would be dramatically modified since in the (near-)zero-temperature limit, the mass would be governed by the chemical potential,
\baa
M = 2^{\frac{d}{2}} \pqty{\frac{d-2}{d-1}}^{\frac{d-2}{2}} \frac{(L\mu)^d}{L^2}.
\ea
Therefore,  the decay time-scale of the quasi-normal modes would be given by $\mu^{- \frac{2d}{d+2}}$,  as anticipated on physical grounds mentioned previously.   
We have verified the scaling with both the momentum and the chemical potential for the fundamental mode $n=0$ in Figure \ref{fig-neutralqnm}.  The close agreement with the predicted results confirms validity of the analysis.

 In the large-$d$ limit,  we find that,
\baa
\omega L = k + \frac{\pi^2 \mu^2 (2n + 1+\nu)^2}{4 k}  \pqty{ 1  - \mi \frac{2\pi}{d} + \cdots}.
\ea

Thus,  in the large $d$ limit,   we have a similar $\mu$ and $k$-dependence on the relaxation time-scale.  It would therefore appear that in the large-$d$ limit,  all quasi-normal modes would have a similar scaling. In fact for an uncharged AdS black brane,  we would have a similar behavior,  with $\mu$ being replaced by $T$. 

\subsubsection{Spherically Symmetric Geometry}\label{ss-spher}

We now return to the black hole geometry which is asymptotic to $\mathbb{R} \times S^{d-1}$.  The effective potential has the form as \eqref{potpla}  with the replacement,
\baa
k^2 \to \qty(\ell + \frac{d -2}{2})^2 -\frac14. \label{repl2}
\ea
When the eikonal limit is now taken,  in complex or real spin,  it is always in the combination,
\baa
\ell + \frac{d -2}{2} = \La \kh, \quad \omega = \La \oh ,  \quad q = \La \qh. \label{jwkbrelsp}
\ea
The shape of the potential for a spherically symmetric geometry will have more structure.   However, for appropriate values of $\kh$,  $\oh$ and $\qh$,  there could be one turning point of the eikonal potential as shown in Figure \ref{fig-turning1}.

When we consider the turning point to be arbitrarily close to the asymptotic infinity,  an interesting dimension-dependent phenomenon takes place. For $d = 3$,   the first term on the right of \eqref{ort} is still the first sub-leading term at large $r_T$.  Therefore,  even in the spherically symmetric situation,  for $qQ < 0$,  eq. \eqref{rtapprox} continues to give the approximate turning point.  Therefore,  eq.  \eqref{qnm1} with $d = 3$ gives the quasi-normal mode for large and small black holes alike.

The case of $d=4$ is special and needs a separate treatment.  In this case,  provided that $\sqrt{3} \qh Q / 2 + \kh L /2  < 0$
\baa
 r_T =\sqrt{ \frac{ - \pqty{\sqrt{3} \qh  Q /2 + \kh L /2 } }{ \frac{\kh}{L}  -  (\oh + \qh \mu) } }.
\ea
This eventually yields the quasi-normal modes
\baa
(\om + \qh \mu) L = k - \mi (2n + 1 + \nu) \sqrt{ - \frac{k + \sqrt{3} q Q / L}{k}}.
\ea

In $d \geq 5$,  there are no real turning points of the eikonal potential arbitrarily close to the boundary.  We can still have complex turning points and the same comments as in the positive charged case,  made towards the end of \S\ref{ss-planar}  apply. When the probe charge $q = 0$,  then there are no turning points in the asymptotic region for all $d \geq 3$.

\subsection{Two Turning Points}\label{ss-twotp}

We now explore the situation where there are two turning points corresponding to the presence of an unstable photon sphere.  A typical form of the potential is shown in Figure \ref{fig-turning2}.   In the radial coordinate $r$,  we call the turning point near the boundary $r_{T_1}$ and the turning point near the horizon $r_{T_2}$.  Note that such trajectories can arise for charged fields in both spherical and planar geometries and for neutral fields in spherically symmetric geometries,  as in \cite{Hashimoto:2023buz,  Dodelson:2023nnr}.  For the sake of completeness and to illustrate the differences in our $\mi \e$-prescription,  let us derive the QNMs.

The region close to the horizon is classically allowed and we can write down the same ingoing solution \eqref{JWKBpal} as before,  with $z_T \to z_{T_2}$ and making suitable modifications in the function $\kappa(z)$ taking into account the geometry in question.   Assuming that the turning points are well separated,  the JWKB solution would be given by
\baa
\psi (z) &= \frac{1}{\sqrt{\Lambda | \kappa (z) |} } \bqty{ A \exp(\La \int_z^{z_{T_2}} \dd{z} |\kappa(z) |   )+ B \exp(- \La \int_z^{z_{T_2} } \dd{z} |\kappa(z) |   )  }\\
&= \frac{1}{\sqrt{\Lambda | \kappa (z) |} } \bqty{ A e^\eta \exp(-\La \int_{z_{T_1}}^{z} \dd{z} |\kappa(z) |   )+ B e^{-\eta} \exp(\La \int_{z_{T_1}}^{z} \dd{z} |\kappa(z) |   )  },  \label{JWKBfr2t}
\ea
where
\baa
\eta \equiv \La \int_{z_{T_1}}^{z_{T_2} } \dd{z} |\kappa(z) | .  \label{etadef}
\ea
The first line of \eqref{JWKBfr2t} is suitable for matching across the turning point near the horizon and the second line is suitable for matching across the turning point towards the boundary. The region $z < z_{T_1}$ is classically allowed and the JWKB solution  is given by
\baa
\psi (z) =  \frac{1}{\sqrt{\Lambda \kappa (z)} } \bqty{ C \exp(\mi \La \int_z^{z_{T_1}} \dd{z} \kappa(z)   )+ D \exp(- \mi  \La \int_z^{z_{T_2} } \dd{z} \kappa(z)   )  }.  \label{JWKBall2p}
\ea

The connection problem at $z =  z_{T_2}$ remains unchanged in comparison with the case of one turning point in \S\ref{ss-planar}. Thus we still have $A = \exp(-\mi \pi /4) = -\mi B$.  The connection formula at $z = z_{T_1}$ gives the connection coefficients
\baa
C = - \mi \sinh \eta ,  \quad D =  \cosh \eta.  \label{conncd}
\ea
Note that $\rho^2$ as defined in \eqref{defrho} is now negative and we can find the asymptotic solution in terms of Hankel functions.  The mode in \eqref{JWKBall2p} associated with $C$ ($D$) gets mapped to the Hankel function of the second (first) kind. We obtain the solution
\baa
\sqrt{\frac{2}{\pi}}\psi (z) = \sqrt{z}  e^{- \mi \pqty{ \theta - \frac{\pi \nu}{2} - \frac{\pi}{4}  }  }  \cosh\eta  \, H^{(1)}_\nu (\La  |\rho| z) - \mi \sqrt{z}  e^{ \mi \pqty{ \theta - \frac{\pi \nu}{2} - \frac{\pi}{4}  }  }  \sinh\eta  \, H^{(2)}_\nu (\La |\rho| z), \label{hanksol}
\ea
where $\theta$,  as before is the integral from the boundary to the first turning point,
\baa
\theta \equiv \La \int_0^{z_{T_1}} \dd{z} \kappa (z).
\ea
 
We can read off the retarded Green function from this expression to be
\baa
G_R (\omega, \ell ) = \pqty{ - \frac{\La^2 \rho^2}{4}  }^\nu \frac{\Gamma(-\nu)}{\Gamma(\nu)} \frac{ e^{2 \mi \theta}  + e^{- \mi \pi \nu} \coth \eta  }{e^{2 \mi  \theta - \mi \pi \nu  }   + \coth \eta }.  \label{rgfrsp}
\ea
The quasi-normal poles are given by,
\baa
\theta = \frac{\pi}{2} ( 2n + \nu + 1) - \frac{\mi}{2} \log ( \coth \eta).   \label{tespresp}
\ea
Expanding for large $\eta$,   so that the turning point close to boundary, we obtain,
\baa
\omega + q \mu = \Delta_+ + \ell + 2n - \frac{2\mi}{\pi} e^{-2\eta} + \cdots.  \label{omspres}
\ea
We thus find, in consonance with \cite{Festuccia:2008zx},  \cite{Dodelson:2023nnr} that there exist long-lived modes. with exponentially long lifetimes.  It is however worth remarking here that we find a minor difference  in the imaginary part of \eqref{omspres} with respect to the results of \cite{Dodelson:2023nnr}.  

{ It is worth elaborating on this discrepancy which we believe stems from our $\mi \e$ prescription.  In matching the JWKB wave functions,  we use the following asymptotic expansions of the Airy functions \cite{Dingle:1973}:
\baa
\mathrm{Ai} (z) &= z^{ - \frac14 } E_-,  \qquad | \arg z | < \frac{2}{3} \pi, \\
\mathrm{Bi} (z) &=2z^{ - \frac14 } E_+  + z^{ - \frac14 } E_- \times   \begin{cases} \mi, \qquad &\pqty{0< \arg z < \frac{2}{3}\pi}, \\ -\mi ,  \qquad &\pqty{- \frac{2}{3}\pi < \arg z <0},\\ 0,  \qquad & \pqty{\arg z = 0},\end{cases} \label{airyasym}
\ea
with
\baa
E_\pm (z) = \frac{\exp( \pm \frac{2}{3} z^{ \frac{3}{2}} )}{2 \sqrt{\pi} \Gamma \qty( \frac{5}{6}  ) \Gamma \qty( \frac{1}{6}  ) } \sum_{r = 0}^\infty \frac{ \Gamma \qty( r+ \frac{5}{6}  ) \Gamma \qty(r +  \frac{1}{6}  )}{\Gamma(r+1) \pqty{ \pm \frac{4}{3}  z^{ \frac{3}{2} } }^r } . \label{epmdef}
\ea
We see from \eqref{airyasym} that the asymptotic expansion for the Airy function $\mathrm{Bi} (z)$ is crucially dependent on $\arg z$.   Our $\mi \e$-prescription necessarily requires a non-zero value of $\arg z$.  In contrast,  for this particular problem,  the authors of \cite{Dodelson:2023nnr} appear to have used $\epsilon = 0$, for which the asymptotic expansion is different.  However,  we emphasize that this disagreement is actually a minor one for large $\eta$ since this would mean at most a sub-leading difference of $\log 2$ in the exponent.  For small values of $\eta$,  the JWKB approximation would break down anyway and we would not be able to use the formula above.   It would be interesting to examine the case where $\eta$ is of moderate values so that the JWKB approximation is valid and one can compute the imaginary part reliably and compare with numerical results.  We hope to come back to this problem in a future work.}

\subsubsection{Complex Spin and Charge}\label{ss-complexcharge}

Working with an asymptotically flat spacetime,   the authors of \cite{Cardoso:2008bp} had derived an analytic expression for quasi-normal modes with wave functions localized close to the photon sphere.   The authors of \cite{Cardoso:2008bp} had noted that the formula derived would not be valid for asymptotically AdS spacetime -- which is indeed true,  as we saw in \eqref{omspres}. One of the remarkable insights provided by \cite{Dodelson:2023nnr} is that the formula, which we discuss shortly, would be valid even in asymptotically AdS spacetime provided we consider a \emph{complex} spin.  For the case of charged particles, we have to take the following limit,
\baa
\ell + \frac{d -2}{2} = \mi \La \kh, \quad \omega = \mi \La \oh ,  \quad q =  \mi  \La\qh. \label{jwkbrelsp2}
\ea
which rotates all the eikonal parameters by the same phase.  It therefore changes the eikonal potential $\kappa^2(z)$ by an overall sign. We could of course choose to rotate only $\kh$ and leaving $\qh$ and $\oh$ as they are.  But the structure of such an eikonal potential would be quite different from the one associated with the photon sphere since such analytic continuation would introduce \emph{relative} signs.  In this context,  taking an imaginary charge is quite a natural thing to do, especially when considering Kaluza-Klein reduction of an imaginary spin mode in higher dimensions. The eikonal potential is now given by,
\baa
\kappa^2 (z) = f(r) \frac{\kh^2}{r^2} - (\oh + \qh A_t )^2 \equiv -\hat{V} (r). \label{eikim2}
\ea
The potential $\hat{V} (r)$ is by construction,  negative of the form of the potential given in \eqref{vrdef}.  The expressions for the values of the frequency $\oh= \oh_0$ and the momentum $\kh = \kh_0$ which satisfies $\hat{V} (r_p ) = 0 = \hat{V} ' (r_p)$ can be read off from eq. \eqref{el2ch} with appropriate replacements of symbols.  Keeping the momentum $\kh$ fixed at $\kh = \kh_0$,  we consider changing the energy slightly so that $\omega = \omega_0 - \delta \omega$ so that we can treat the modes in a harmonic oscillator approximation.  We can thus use the standard JWKB quantization condition for a bounded semiclassical trajectory (also known as the Bohr-Sommerfeld quantization condition)
\baa
\La \int_{z_{T_1}}^{z_{T_2}} \dd{z} \kappa (z) = \pqty{ n + \frac{1}{2}  } \pi.
\ea
On applying this formula, we find, in the end
\baa
\om = - q A_t (r_p ) + \mi  k_0 \sqrt{\frac{f(r_p)}{r_p^2}} - \mi  \pqty{ n + \frac{1}{2}}  \sqrt{ \frac{\hat{V}''(r_p) f(r_p) r_p^2 }{2\kh_0^2} }. \label{finalans}
\ea
The final square-root term above is referred to as the Lyapunov exponent \cite{Cardoso:2008bp} for good reason.  As explained for example in \cite{Hadar:2022xag}, this exponent controls the windings of photon orbits close to the photon sphere.

Since we are interested in taking $r_p$ close to the extremal horizon,  writing $q =\mi q_0$,
\baa
\om &= \mi (-q_0 + |q_0| ) \sqrt{ \frac{(d-1)(d-2)}{2} } \sqrt{\frac{1}{r_h^2} + \frac{d}{d-2} \frac{1}{L^2} } (r_p - r_h) \\
&\quad -  \mi  \pqty{ n + \frac{1}{2}}  \sqrt{ \frac{2 [ d(d-1) r_h^2 + (d-2)^2 L^2 ] [ 2 d (d-1) r_h^2 + 3(d-2)^2 L^2 ] }{3L^4 r_h^5}  (r_p - r_h) }. \label{omecomspch}
\ea
For negative $q_0$,  the leading term is positive and we would have the QNMs starting in the upper-half-plane.  For a positive $q_0$,  a term at the next order would be important,  and it will still be on the upper-half-plane,  see eq.  \eqref{elexp}.   We have attempted a numerical verification of the results in this subsection --- but no reliable results in physically interesting regimes could be obtained by use of the aforementioned package.  We hope to come back to a detailed numerical study of such QNMs in the future.

\section{Superradiant Modes in the Eikonal Limit}\label{sec-superr}

We have already seen the emergence of several interesting features of charged fields,  in contrast with neutral fields.  Let us now consider another distinct phenomenon involving charged fields that has no analog for neutral fields in spherically symmetric backgrounds.   This phenomenon goes by the name of \emph{superradiance},  which is extensively studied in the black hole literature \cite{Brito:2015oca}.  If the wave vector of the ingoing mode on the black hole horizon is directed away from the event horizon,  there is a superradiant amplification of the impinging wave.  In other words,  superradiance occurs when the asymptotic frequency, $\om + q \mu > 0$,  but the frequency on the black hole horizon, $\om < 0$.  This necessarily implies that $qQ > 0$.

The method of analytic continuation proposed in this article can be readily extended to include superradiant modes as well.  The JWKB solution which is ingoing on the horizon is now described by the exponential of the other sign (in contrast with eq. \eqref{JWKBpal})
\baa
\psi (z) = \frac{1}{\sqrt{\Lambda \kappa (z) }} \exp( -\mi \La \int_{z_T}^{z} \dd{z}\kappa (z)    ),  \qquad (z > z_T). \label{JWKBpal2}
\ea
This tells us that the appropriate deformation of $\Lambda$ must now be in the \emph{lower} half-plane,
\baa
\Lambda \to \Lambda (1 - \mi \e). \label{iep2}
\ea
The JWKB wave function, still given by \eqref{JWKBpd}, will now have different connection coefficients,  $A = \exp(\mi \pi /4) = \mi B$. 

When one considers the situation with only one turning point as in \S\ref{ss-onetp}.  In terms of the modified Bessel functions,  the asymptotic solution would then be given by
\baa
e^{ - \frac{\mi \pi}{4} } \psi(z) = - \sqrt{2\pi} \mi e^{- \theta}  \sqrt{z} I_\nu (\Lambda \rho z) + \sqrt{\frac{2}{\pi}} (e^\theta + e^{-\mi \pi \nu} e^{-\theta} ) \sqrt{z}   K_\nu (\Lambda \rho z),  \label{besol2}
\ea
with the corresponding retarded Green function
\baa
G_R (\omega, k) =  \frac{\Gamma(-\nu)}{\Gamma(\nu)}  \pqty{ \frac{\Lambda^2 \rho^2}{4} }^\nu e^{\mi \pi \nu} \frac{\cosh( \theta - \frac{\mi \pi \nu}{2}  )}{\cosh( \theta + \frac{\mi \pi \nu}{2}  )}. \label{greta2}
\ea
The quasi-normal modes would then be given by
\baa
\theta = - \frac{\mi \pi}{2} (2n + 1 + \nu),  \quad n \in \mathbb{Z}_{ \geq 0}. \label{quancon2}
\ea
The correct sign for the \eqref{quancon2} can be ascertained by the same $\mi \epsilon$-prescription as in \eqref{iep2}.  As mentioned at the end of \S\ref{ss-planar},  for $qQ > 0$,  there does not exist a real turning point for arbitrarily large values of the radial coordinate $r$.  A similar na\"ive argument with the phase of the charge being rotated now suggests that the quasi-normal poles would be located on the upper-half $\omega$-plane, signifying an instability.

This is much clearer to illustrate for the case involving two turning points as in \S\ref{ss-twotp}. In this case, the analysis proceeds similarly with a suitable modification of the connection coefficients $A$ and $B$ in the classically forbidden region,  \eqref{JWKBfr2t}.  The JWKB wave function in the classically allowed region near the boundary is given by \eqref{JWKBall2p}.  On account of the new connection coefficients $A$ and $B$,  the connection coefficients $C$ and $D$ are given by
\baa
C = \cosh \eta ,  \quad D = \mi \sinh \eta \label{conncd2},
\ea
with a corresponding solution in terms of Hankel functions as in eq. \eqref{hanksol}.
The retarded Green function in this case is given by
\baa
G_R (\omega, \ell ) = \pqty{ - \frac{\La^2 \rho^2}{4}  }^\nu \frac{\Gamma(-\nu)}{\Gamma(\nu)} \frac{ e^{2 \mi \theta}  + e^{- \mi \pi \nu} \tanh \eta  }{e^{2 \mi  \theta - \mi \pi \nu  }   + \tanh \eta }.  \label{rgfrsp2}
\ea
The quasi-normal poles are now located at
\baa
\theta = \frac{\pi}{2} (2n + 1 + \nu ) - \frac{\mi}{2} \log( \tanh \eta). \label{srJWKB2}
\ea
For large $\eta$,  the difference between \eqref{srJWKB2} and \eqref{omspres} would appear to be inconsequential, but the $\tanh$ function makes all the difference in the world. The imaginary part of $\omega$ is now exponentially small but \emph{positive},
\baa
\omega + q \mu = \Delta_+ + \ell + 2n + \frac{2\mi}{\pi} \exp(-2\eta).
\ea
Therefore,  such superradiant perturbations are manifestly \emph{unstable} and destabilize the geometry. The time-scale of the instability is exponentially large in $\eta$,  $e^{2 \eta}$.  

\begin{figure}
\centering
\includegraphics[width=12cm]{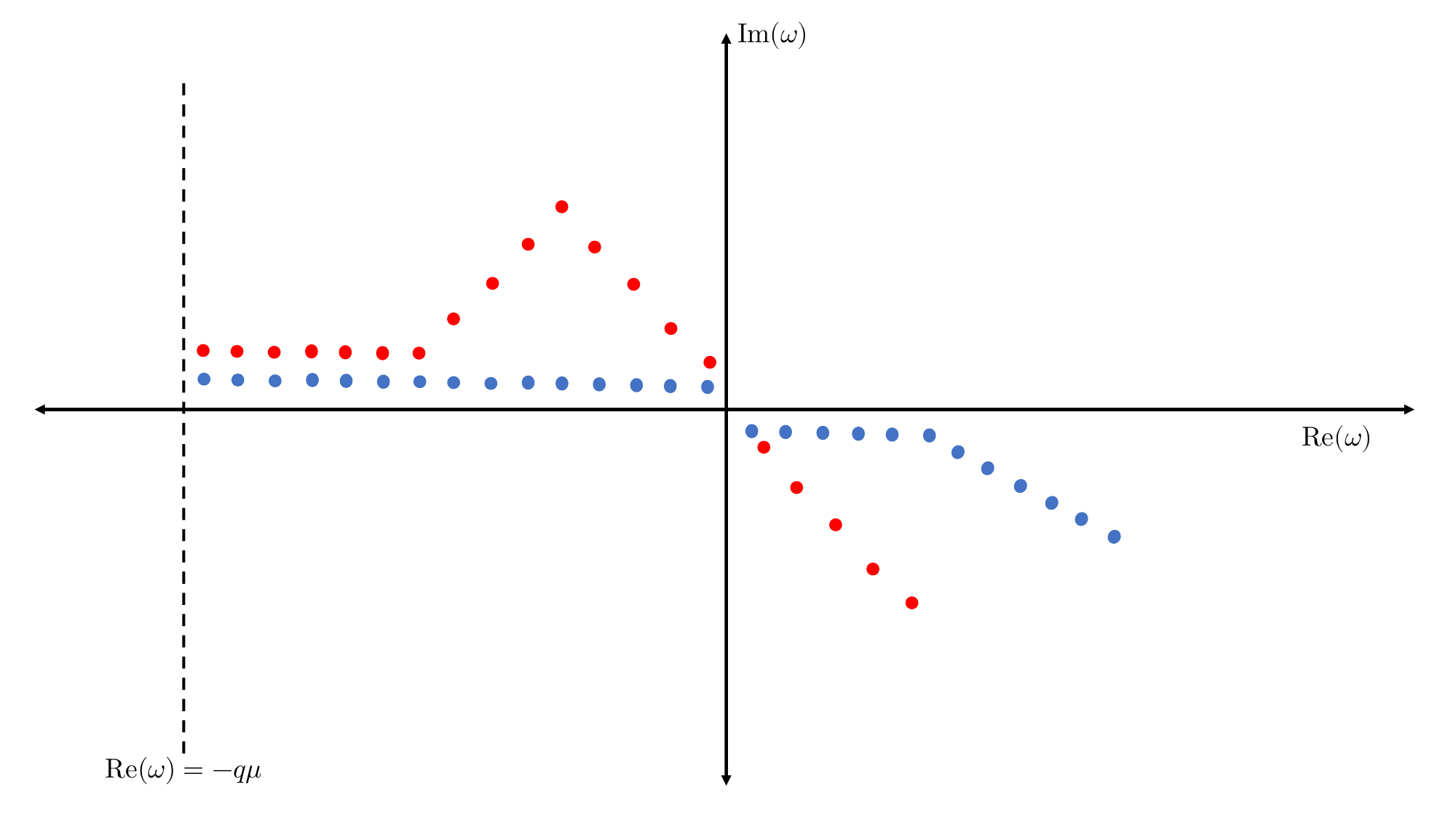}
\caption{A schematic depiction of two possible behaviors of superradiant QNMs on the complex $\omega$-plane.}
\label{fig-super}
\end{figure}

The quasi-normal poles will have different behavior depending on the values of $\Re(\omega)$,  $\ell$ and $q \mu$. For small overtone number $n$ for which the photon ring peak is a classically forbidden region,  the poles would lie on the upper half-plane close to the real axis.  For higher overtone numbers $n$  if $\Re (\omega) < 0$ even after the particle is highly energetic,  then superradiant instability continues to affect the geometry beyond this value of $n$ and the positive imaginary parts would be no longer exponentially suppressed.  At very high values of $n$,  when eventually $\Re(\omega) > 0$, superradiance stops and the poles move into the lower half-plane.

 On the other hand, if superradiance stops even before the particle reaching the top of the charged photon potential,  the exponentially small positive parts move on to exponentially small negative parts,  a pattern that persists until the classical path reaches the top of the photon sphere.  The two qualitatively different scenarios are shown in Figure \ref{fig-super},  compare with \cite{Hashimoto:2023buz,  Dodelson:2023nnr}. Note that throughout this work,  we have considered the quadrant corresponding to $\Re ( \omega + q \mu ) > 0$.  The form of the quasi-normal poles on the other quadrant $\Re ( \omega + q \mu ) < 0$ will in general not be symmetric about $\Re ( \omega + q \mu ) = 0$ unlike in the uncharged case.  The nature of the poles in these cases can be ascertained by a generalization of the results in this paper.

\section{Discussion}\label{sec-disc}

One of the most immediate consequences of the results of the paper is that many of the calculations in \cite{Dodelson:2023nnr} pertaining to bulk-cone singularities,  done for non-extremal black holes,  would go through almost unchanged even when a near-extremal geometry is considered. As mentioned previously,  in many cases of interest, the chemical potential replaces the temperature as the dominant energy scale.  Our results also apply to near-extremal rotating black holes in higher-dimensional spacetimes upon dimensional oxidation. The eikonal limit is quite general --- while we have considered a free scalar field,  the eikonal limits of gravitational and electromagnetic perturbations reduce to the same form.  Therefore,  the behavior of the corresponding operators on the CFT side will be captured by the results here.   The results here would also be pertinent to large-charge CFTs \cite{Hellerman:2015nra} when there is a large ground state degeneracy.  It will be interesting to ponder a bulk dual without black holes \cite{Loukas:2018zjh}.

One of the main motivations for this work was to explore the possibility of bringing the photon ring to the region of spacetime well-described by the JT model.  This has been accomplished in no fewer than two different ways, involving large dimensions and charged particles.  {Describing gravitational dynamics within the relatively simple JT model typically endows one with a remarkable degree of analytical control which is otherwise elusive,  often involving complex computations in higher dimensions.   One of the most appealing features of the JT model is its \emph{universality} in  describing the dynamics of nearly extremal black holes irrespective of the details of the asymptotic nature of spacetime geometry \cite{Moitra:2019bub,  Moitra:2022ooc}.  Describing the light rings within the framework of the JT model would amount to an simple effective theory of matter fields coupled to Schwarzian theory and the phase mode \cite{Moitra:2018jqs}.   It will also be interesting to consider the question directly in the charged versions of the Sachdev-Ye-Kitaev model \cite{Davison:2016ngz, Gu:2019jub}.  A low-dimensional effective theory will also be able to shed light on features of photon rings in other generic near-extremal backgrounds.   The JT model path integral is useful for studying non-negligible quantum corrections \cite{Preskill:1991tb} in near-extremal geometries as well. }  On account of the near-horizon symmetries,  such corrections are often under good computational control even in the absence of any supersymmetry. For instance,  even at small enough temperatures $T \sim \mathcal{J}/S_0$,  one can very easily obtain the $T^{3/2}$ correction factor in front of the partition function for general black hole backgrounds by combining the results of \cite{Stanford:2017thb} and \cite{Moitra:2019bub}. Even in this case,  JT gravity continues to be a good description of the geometry.  In fact,  pertinent to the picture presented in this article,  the effects of additional matter fields can also be incorporated \cite{Moitra:2021uiv} in the path integral in a straightforward manner, possibly with some renormalization to account for the high-energy modes. 

It would be quite fruitful to consider other numerical techniques to gain a better understanding of the gravitational dynamics governing the perturbations.   A useful application of such techniques might shed light even on dynamical spacetimes of the kind considered in \cite{Moitra:2022umq}.  We look forward to exploring these exciting possibilities in future works\footnote{In other words, this is where one has to say ``Nox'', for now.}.

\acknowledgments

I would like to thank the following organizations for their warm hospitality at various stages of this work: University of Amsterdam (during the String Summer Workshop 2024),  University of California Santa Barbara and the Simons Center for Geometry and Physics, Stony Brook University (during the Simons Physics Summer Workshop 2024).  
Support for the final stage (revision and publication) of this work was provided by the European Research Council under the European Union's Seventh Framework Programme (FP7/2007-2013), ERC Grant agreement ADG 834878.
 \bibliographystyle{JHEP}
 \bibliography{lumos_ref}

\end{document}